# Are Generics and Negativity about Social Groups Common on Social Media? – A Comparative Analysis of Twitter (X) Data

Uwe Peters (Dept. of Philosophy, Utrecht University)
Ignacio Ojea Quintana (Munich Center for Mathematical Philosophy, LMU) *

## Abstract

Many philosophers hold that generics (i.e., unquantified generalizations) are pervasive in communication and that when they are about social groups, this may offend and polarize people because generics gloss over variations between individuals. Generics about social groups might be particularly common on Twitter (X). This remains unexplored, however. Using machine learning (ML) techniques, we therefore developed an automatic classifier for social generics, applied it to 1.1 million tweets about people, and analyzed the tweets. While it is often suggested that generics are ubiquitous in everyday communication, we found that most tweets (78%) about people contained no generics. However, tweets with generics received more "likes" and retweets. Furthermore, while recent psychological research may lead to the prediction that tweets with generics about political groups are more common than tweets with generics about ethnic groups, we found the opposite. However, consistent with recent claims that political animosity is less constrained by social norms than animosity against gender and ethnic groups, negative tweets with generics about political groups were significantly more prevalent and retweeted than negative tweets about ethnic groups. Our study provides the first ML-based insights into the use and impact of social generics on Twitter.

## 1. Introduction

People often reject the way they are labelled or categorized by others because they dislike how others view them (Hacking, 1995). One way in which people categorize each other is by using *generics* (Leslie, 2017; Gelman, 2021). Generics are generalizing sentences that make broad claims about a whole category of individuals (people, things, etc.) without a quantifier (e.g., 'many', 'some', '75%') in the noun phrase that is describing the subject of the sentence (Krifka et al., 1995). Generics ascribe features to entire classes of individuals (e.g., 'students like to party', 'children need supervision', 'liberals favor equality'), not specific subsets of them. They are thought to be the "most common" and "most

---

fundamental" way in which humans generalize (Leslie, 2017; DeJesus et al., 2019, p. 18371).

While generics often have other referents than people (Carlson & Pelletier, 1995), the focus here will be on generics about people or social groups (henceforth 'social generics'). These kinds of generics are particularly normatively interesting since they may be especially likely to fuel societal problems. This is because generics obscure variations within social groups by depicting all members of a group as alike with respect to an ascribed feature (e.g., 'girls like pink', 'Muslims are terrorists') (Leslie, 2017). Relatedly, generics about people have a "power to offend" and divide by eliciting aversive responding from group members who may feel misrepresented by the sweeping claims that generics may convey (Gelman, 2021, p. 517). Given these problematic features of generics, it is important to investigate where and how frequently they are being used in communication.

While there are different domains of communication where one may study social generics, worldwide about 5.04 billion people (almost 62.3% of the global population) now use social media platforms such as Twitter (now X),[1] Facebook, Instagram, or TikTok to communicate (Petrosyan, 2024). Twitter is particularly attractive for studying social generics because of its relatively easy data access interface (Antonakaki et al., 2021) and because for most users (i.e., those not paying for the platform), Twitter only provides limited space for posts (280 characters; Weatherbed, 2023), which may increase the use of generics as they are shorter than quantified generalizations. Additionally, Twitter is known to incentivize the tweeting of enraging, attention grabbing, bold claims (Rose-Stockwell, 2023), a feat that generics can, due to their broad scope, facilitate. Exploring the distribution of generics on Twitter is therefore especially relevant because the platform's design and algorithms may drastically increase people's use and exposure to generics, potentially exacerbating social polarization online by promoting exaggerated depictions of between-group differences (Kramer et al., 2021).

Social polarization is currently in many countries particularly prevalent in the *political* domain (Westwood et al., 2018; Reiljan, 2021). Since Twitter is often viewed as a key contributor to political polarization, i.e., the affective or ideological division and aversion between people of different political orientations (Hong & Kim, 2016; Conover et al., 2021), investigating the use of generics about *political* groups on Twitter becomes especially interesting. For instance, several researchers have argued that while there are strong social norms that put constraints on expressions of hostility, stereotypes, and bias against gender or ethnic groups, there are no similarly strong norms against them regarding political opponents (Iyengar & Westwood, 2015; Peters, 2022). One may thus predict that on Twitter the use of generics about political groups is much higher than the use of generics about gender or ethnic groups because generics can effectively communicate stereotypes and dichotomous thinking about social groups (Leslie, 2017).

---

[1] We are using the company's old name here because at the time of preregistration of this study, Twitter had not yet changed its name. Our pre-registered material and hypotheses still contain the old name. For consistency, we therefore retain it.

However, while several observational studies have investigated the distribution of generics in, for instance, academic (including philosophical) texts (DeJesus et al., 2019; Peters & Lemeire, 2023), and generics are a hot philosophical topic (e.g., Leslie & Lerner, 2022; Chin-Yee, 2023; Stegenga, 2024), the use of generics about social groups in Twitter communication remains unexplored in the sciences and philosophy. Moreover, the existing observational studies of academic texts rely on manually classifying and extracting generics from text corpora, leaving significant room for human error in the data extraction. A large-scale investigation of generics use on Twitter would also involve big data processing of potentially millions of data points, requiring an automatic classification tool. However, although automatic classifiers for generics have been developed (Ralethe & Buys, 2022), no such tool specifically for *social* generics exists yet. The scientific study of the perhaps most fundamental and potentially most problematic way in which we categorize people on social media is therefore currently severely limited in scope.

To begin filling these research lacunas, we developed (and made freely available here) a prototype supervised machine learning model for the classification of social generics on Twitter. We applied this model to more than 1 million tweets to analyze the use and impact of social generics on Twitter, comparing generics about political, gender, and ethnic groups. In the next sections, we elaborate on the background of the project, specify our study hypotheses, and outline our methodology before reporting the analysis results and discussing their implications.

## 2. Background and related research

Generics are of significant philosophical and scientific interest. For instance, while generics are present across all languages, they have puzzling truth conditions (Carlson & Pelletier, 1995): Some generics are true even if 90% of group members lack the property that the generics ascribe to them (e.g., 'mosquitoes carry malaria') while others remain false even if most group members have the ascribed characteristic (e.g., 'books are paperbacks') (Leslie & Lerner, 2022). Yet, despite the semantic complexity of generics, psychologists found that children could understand them earlier than claims with quantifiers ('all', 'some', 'most'; Hollander et al., 2002). Moreover, in experiments, people accepted generics based on little evidence about only some members of a group despite interpreting them as applying to all members (Cimpian et al., 2010). Quantified statements were also frequently misremembered as generics, suggesting that human cognition may have a bias towards generic generalizations (Sutherland et al., 2015). Additionally, corpus analyses found that even in scientific articles, where one may expect to find qualified generalizations, generics were used in about 70–90% of articles (DeJesus et al., 2019; Peters & Lemeire, 2023). Indeed, generics are thought to be people's "default mode" of generalization (Leslie, 2012) and be "ubiquitous in everyday communication, thought, and scientific discourse" (Sterken, 2016, p. 17; Liebesman, 2011, p. 409).

The use of generics can be problematic, however. By referring to whole categories of individuals without a quantifier (e.g., 'girls like pink'), generics do not only downplay variability but also communicate only vague prevalence levels, making them harder to assess than (e.g.) universally quantified claims ('all *K*s') (Tessler & Goodman, 2019). The generic '*K*s are *F*' might mean, for instance, 'some', 'many', 'most', '75%', or 'all *K*s are *F*'. People therefore need more background information to determine what a generic conveys than what a quantified generalization conveys, leaving more room for reasoning fallacies and "inferential slippages" to broader claims than warranted (Stegenga, 2024), which can facilitate misinterpretations causing real world harm (e.g., in science communication or clinical contexts; Peters, 2020; Chin-Yee, 2023). Moreover, generics suggest that a particular feature may be conceptually central to the identity of members of a group, encouraging essentialist thinking: A generic such as 'women wear dresses' can license stronger inferences about women than the statement '*these* women wear dresses' since it suggests the property is an inherent one that generalizes across category members (Roberts, 2022). Relatedly, many social generics are viewed as linguistic manifestations of gender and ethnic stereotypes that can be difficult to reject because unlike universal quantifications, generics permit exceptions (Leslie, 2017) (e.g., the generic 'the Russians are violating Ukrainian sovereignty' is not disproven by mentioning some Russian civilians not involved in violating Ukrainian sovereignty). Proponents of generics may thus (if challenged) shift to weaker interpretations of their claims (Lemeire, 2021). Finally, by obscuring variation between individuals, social generics may exacerbate social tension more than quantified claims, which can be more easily disproven or explicitly do not apply to all group members. In fact, some researchers have suggested that social generics may be so problematic that people should avoid them (Leslie, 2017). Yet, others have highlighted their potential benefits (Ritchie, 2019). While there is an ongoing debate on the costs and benefits of using social generics, it is a common view that since generics "essentialize" and "stereotype", their apparent "pervasiveness" is "troubling" (Gelman, 2021, p. 528).

It remains unknown, however, to what extent social generics are also being used on Twitter. Some software designers have developed models that can classify generics in social media data. For instance, Ralethe and Buys (2022) used the base versions of BERT and RoBERTa as pre-trained language models to fine-tune them for the classification of statements as generics. Similarly, Allaway et al. (2023) developed a generics classifier that also enumerates exceptions to generics. However, for their algorithms, Ralethe and Buys and Allaway et al. only used generics about non-human animals, not people. However, given the just mentioned problematic features of generics about people, developing an algorithm that helps gain insight into the distribution and impact of such generics on Twitter may be especially important.

### 3. Hypotheses

We now introduce and motivate five hypotheses that capture the points about generics we have made so far and that an automatic social generics classifier could help test.

To begin with, many commentators report that generalizations about people are "all over social media, Twitter especially" (Edwards, 2017; Twist, 2018; Lister, 2022). This seems unsurprising, as "group identities are hypersalient on social media, especially in the context of online political or moral discussions" (Rathje et al., 2021), and research found that, even offline, "gossiping" about people including generalizations about groups accounted for about 65% of speaking time (across age or gender), as it strenghtens group ties by establishing in-group/out-group boundaries and boosting self-esteem (via enabling downward comparisons) (Stambor, 2006). Social media posts that generalize about (e.g., political) outgroups have also been found to receive high engagement scores (Rathje et al., 2021). Adding to these considerations the point that it is frequently claimed that generalizations, more specifically, generics are pervasive in language use in general (Table 1 presents textual evidence of such claims), one may predict that social generics are also highly common in tweets, especially given the fact that for most users, Twitter imposes strict textual space constraints (Weatherbed, 2023).

| *How common are generalizations and generics?* |
| --- |
| (1) The "language of generalization is ubiquitous in everyday conversation". (Tessler & Goodman, 2019, p. 4) |
| (2) "Generic sentences are ubiquitous". (Liebesman, 2011, p. 409) |
| (3) "Generics are ubiquitous". (Meyer et al., 2011, p. 913) |
| (4) "Generics [are] ubiquitous in everyday communication, thought, and scientific discourse". (Sterken, 2016, p. 17) |
| (5) Generics are people's "most fundamental" way of generalizing. (DeJesus et al., 2019, p. 18371) |
| (6) "Generic statements are pervasive." (Reuter et al., 2023, p. 1) |

**Table 1.** Claims about the pervasiveness of generalizations and generics.

Combining these points, the following first hypothesis arises:

*H1: On Twitter, tweets (about people) that contain social generics are more common than tweets (about people) that do not contain social generics.*

Relatedly, it is often assumed that exaggerated, overgeneralizing language ("hype") may draw more attention to claims, helping to underline their importance (Sumner et al., 2016; Intemann, 2022; Peters et al., 2022). Since social generics are, due to their broad scope, apt vehicles for claims with these features (DeJesus et al., 2019) and many Twitter users aim to attract attention (Sherman et al., 2018), one may predict that tweets with social generics are also more impactful than tweets without them. Two tweet impact proxies are "likes" and retweets, where more "likes" or retweets mean higher impact (Lahuerta-Otero et al., 2018). A "like" indicates that a person appreciates or agrees with a tweet. For a retweet, this may not always be the case. Many Twitter profiles state "Retweeting does not mean endorsement." However, surveys found that most participants used retweets, just as "likes", to express endorsement, specifically, a stronger endorsement than that underlying "likes", as retweeted tweets will also feature on the retweeters' own profile and be broadcast to their followers (Metaxas et al., 2015), which can play important online community-building roles by helping boost signals of allies (Ojea Quintana et al., 2022). Based on these points, we predict:

*H2: Tweets with social generics have higher "likes" and retweet impact than tweets without them.*

Furthermore, since studies suggest that political polarization may be common on Twitter especially in Western countries (e.g., the US, UK) (Hong & Kim, 2016) and generics can facilitate and manifest antagonistic thinking and labelling (Kramer et al., 2021), Twitter users may often employ generics in tweets to derogate political opponents or distinguish their own political group. Indeed, comparative studies involving the implicit association task found that participants had stronger political biases than race biases (Iyengar & Westwood, 2015), leading some researchers to note that "unlike race, gender, and other social divides where group-related attitudes and behaviors are subject to social norms [...], there are [e.g., in the USA] no corresponding pressures to temper disapproval of political opponents" (Iyengar et al., 2019, p. 133). Applying these points to Twitter, since incivility, bias, and stereotyping about political groups (e.g., opponents) seem to be more socially acceptable than stereotyping of gender and ethnic groups (Westwood et al., 2018; Peters, 2022), we predict:

> *H3: The use of tweets with generics about political groups is higher than the use of tweets with generics about gender or ethnic groups.*

Indeed, in Western countries, verbal animosity in the political domain is frequently on display in mainstream news (e.g., a recent UK prime minister called a political opponent a "muttering idiot", Donald Trump tweeted "only a dead Democrat is a good Democrat", etc., Moody-Adams, 2019) and on social media (Frimer et al., 2023). This may shape Twitter exchanges such that people feel less constrained toward, and more frequently use, negatively valenced generics about political groups (e.g., opponents) compared to generics about other groups, leading to the prediction:

> *H4: Negatively valenced tweets with generics about political groups are more common compared to negatively valenced tweets with generics about gender or ethnic groups.*

Moreover, since there is evidence that moral or political outrage is particularly likely to spread online (Crockett, 2017) and social generics facilitate exaggerated descriptions of political opponents, which may boost outrage (by causing offence; Gelman, 2021), tweets with such generics might have especially high impact. For instance, in-group members may use them to signal group alliance (via "likes"), express anger, or provoke responses from out-group targets via tweets (Rose-Stockwell, 2023) (e.g., "Democrats glorify the killing of the unborn", "Conservatives are conspiracy theorists") whereas outgroup members may retweet them to express outrage. A fifth and final hypothesis thus arises:

> *H5: Tweets with generics about political groups have higher "likes" and retweet impact than tweets with generics about gender or ethnic groups.*

To test these five hypotheses, we first registered them on an Open Science Framework (OSF) platform[2] before conducting an extensive analysis of Twitter data.

---

[2] The registration is available <u>here.</u> Feasibility constraints meant we could only test a subset of all registered hypotheses. To improve readability and precision, we also re-ordered and re-phrased the tested ones here. H1 was added after preregistration.

# 4. Methodology

In March 2023, we first collected tweets about social groups by using the Twitter API. We then developed several automatic classifiers that could identify social generics in tweets. After comparing the classifiers' performance, we applied the model with the best performance to all tweets and analyzed them for generics, sentiment, and impact to test our five hypotheses. We now elaborate on these methodological steps.

*Data collection and preparation*

*Twitter Data.* To obtain tweets relevant for our hypotheses, we designed a query for original and quote tweets in English containing multiple variants of nouns and adjectives referring to three key Western social groups, i.e., political, gender, and ethnic groups (e.g., 'liberals', 'conservatives', 'Democrats', 'men', 'women', 'Asians', 'Africans', etc.; for the complete query, see the Supplementary Material).[3] We focused on English tweets and Western groups because political polarization and political animosity are thought to be especially pronounced in anglophone countries and claims in the literature about an asymmetry in social norms regarding political, gender, and ethnic groups focus primarily on Western countries (Westwood et al., 2018; Iyengar et al., 2019; Peters, 2022).

All tweets were obtained using small batches from random times during 2022, resulting in 1,519,821 tweets. 50,000 were used as training data. The remaining 1,469,821 tweets were automatically categorized (via the search query) as being about one of four social groups. 684,526 (46.57%) were categorized as about political groups only, 175,395 (11.9%) as about gender groups only, and 211,344 (14.3%) as about ethnic groups only. 398,556 (27.1%) tweets were about more than one group. To prevent double counting, we removed them, leaving a total of 1,071,265 tweets for analyses.

*Generics identification*. There are no established criteria for identifying social generics in tweets. However, there are studies in which researchers have manually extracted generics from academic texts based on specific criteria. One such study is DeJesus et al. (2019), whose criteria focus on complete academic English sentences. However, tweets routinely deviate from such standard, often containing incorrect grammar, omissions, emojis, or images (Ahmed, 2014; Heraldine & Handayani, 2022). To prevent missing social generics in tweets, we therefore adopted a broader operationalization of generics, construing them as any text or text/image combination that conveys (grammatically or ungrammatically; textually or pictorially) claims about a category of individuals without having a quantifier in the noun phrase.

Correspondingly, we extended DeJesus et al.'s criteria. We first had two researchers with expertise on generics and familiarity with Twitter language examine 500 random tweets and identify recurring linguistic structures that they would view as communicating generics even if these structures were ungrammatical, incomplete (e.g., "Asian people: Smart as hell"), contained emojis or referred to images (e.g., "White people when a minority tells them they can't say a slur:"). We subsequently added these structures to DeJesus et al.'s coding criteria for generics before using them as classification instructions (for details, see the Supplementary Material).

---

[3] Since the search nouns may equally appear in quantified generalizations, our keywords did not bias the data retrieval toward generics.

*Training Data Sets.* To train our models, for the classification task, we used three datasets totalling 8,000 examples. The first contained 2,000 sentences from the Internet (e.g., from the BBC) covering a range of social groups and topics (e.g., politics, jobs, entertainment). The second contained 4,000 random tweets from the 50,000 tweets reserved for training. The third consisted of 2,000 tweets (from the same reserve) that were classified as generics by our prototype models, checked by us, and then re-cycled for training to strengthen classification performance. This fine-tuning made the results more interpretable. All training examples were labelled (generic/non-generic) and cross-checked by a researcher with generics expertise who followed the mentioned generics classification instructions.

## Classification Task

To decide which tweets contained generics, we established a binary classification task (generic vs. non-generic). We developed both logistic regression and pre-trained deep learning models (Bert, DistilBert) (Devlin et al., 2019; Sanh et al., 2019) as well as different embeddings (Bag of Words, and a pre-trained transformer embedding) (Song & Raghunathan, 2020). Embeddings are functions from strings to points in a higher dimensional space. Bag of words send each tweet to a vector counting the tokens of words (in random order). Pre-trained embeddings estimate the conditional probability of a word occurring given the other words in the context (Mikolov et. al., 2013). Different embeddings were used because the gain in performance from the more advanced models does not always justify their computational complexity (Rudin, 2019). For performance metrics, we used the area under the curve (AUC) of the receiver operating characteristic, the binary accuracy, and the F1-score. Table 2 shows the classification task results.

| *Classifier* | *Binary accuracy* | *AUC* | *F1* |
| --- | --- | --- | --- |
| Logistic Regression with Bag of Words | 0.76 | 0.82 | 0.38 |
| Logistic Regression with pre-trained embedding | 0.77 | 0.85 | 0.67 |
| DistilBert | 0.90 | 0.89 | 0.72 |
| Base Bert | 0.92 | 0.87 | 0.70 |

**Table 2.** Performance of the models

After comparing the models, we chose the classifier with the highest overall performance for our analysis, which was a distilled version of Bert (Sanh et al., 2022).[4] This model was used for both a *binary generics classification* based on the default 0.50 decision threshold, and a probabilistic *genericity score*, i.e., a continuous value between 0 and 1 capturing the likelihood that a given tweet contains a social generic. All training data, labelling instructions, and models are available on an OSF platform here.

---

[4] It is worth noting that a simple logistic regression with a bag of word representation performed well, particularly on the AUC (Table 1). This aligns with other research indicating that traditional ML methods may often perform as well as or better than deep learning methods (Bailly et al., 2022).

*Sentiment and impact analyses*

Our hypotheses were about social generics but also about negativity (sentiment) and tweet impact. For the sentiment prediction, we used a pre-trained RoBerta-based model specifically designed for Twitter and fine-tuned for sentiment analysis with the TweetEval benchmark (Barbieri et al., 2020). For the impact analyses, we focused on "like" and retweet counts because they are known to be the most common and strongest indicators of tweet influence (Metaxas et al., 2015; Mastroeni et al., 2023).

*Hypothesis testing*

For the testing of H1–H5, we had five dependent variables. These were three scale variables (genericity score, "likes", retweets), and two categorical variables (binary generics categorization, and sentiment (negative, neutral, positive)). Our main independent variable was a tweet's being about a political, gender, or ethnic group. Since Kolmogorov–Smirnov tests of data normality showed that our data were not normally distributed, we used non-parametric statistics, specifically, Pearson $\chi^2$, Mann-Whitney $U$, and Kruskal-Wallis $H$ tests, which make fewer assumptions about the data, can accommodate even large skews and differences in sample sizes, and are more conservative than parametric tests. We adopted the standard significance threshold of α = 0.05. Analyses were two-tailed and conducted using JASP and IBM SPSS 29 (for reproduction, see the data on our OSF platform here).

## 5. Results

*Descriptive data.* From the total number of 1,071,265 tweets about social groups, 63.9% ($n$ = 684,526) were about political, 19.7% ($n$ = 211,344) about ethnic, and 16.4% ($n$ = 175,395) about gender groups. Moreover, 64.5% ($n$ = 690,565) had negative, 26.0% ($n$ = 278,206) had neutral, and 9.6% ($n$ = 102,494) had positive sentiment. The results for each of our five hypotheses were as follows.

*H1: On Twitter, tweets (about people) that contain social generics are more common than tweets (about people) that do not contain social generics.* To examine H1, we first used a $\chi^2$ test to compare the percentages of the two categorical variables (i.e., tweets with versus tweets without generics) and assess the null hypothesis (H0) that the percentages are equal. We could reject the H0 ($\chi^2$ (1) = 327051.32, $p$ < 0.001). However, the direction of the difference was unlike predicted because tweets with social generics (≥ 0.50 binary categorization threshold) constituted only 22.4% ($n$ = 239,677) of all tweets.

*H2: Tweets with social generics have higher "likes" and retweet impact than tweets without them.* To examine H2, we conducted a Mann-Whitney $U$ test with the binary generics categorization as the independent variable and "like" and retweet counts as dependent variables. The test uses mean ranks (higher ranks correspond to higher means) to assess the H0 that the two groups' average ranks are the same. As predicted, we found that tweets with social generics received more "likes" (*Mean rank* = 541819.27) than tweets without them (*Mean rank* = 533850.02, $U$ = 9.82E+10, $z$ = – 11.68, $p$ < 0.001, $r$ = 0.011). Tweets with social generics also had significantly higher retweet impact (*Mean rank* = 545752.08) compared to tweets without them (*Mean rank* = 532716.52) ($U$ = 9.72E+10, $z$ = – 24.177, $p$ < 0.001, $r$ = 0.023). These results support H2.

*H3: The use of tweets with generics about political groups is higher than the use of tweets with generics about gender or ethnic groups.* Testing H3, we first found that in absolute terms, of all tweets with social generics (binary categorization) ($n$ = 239,677), 60.4% ($n$ = 144,789) were about political, 26.3% ($n$ = 63,042) about ethnic, and 13.3% ($n$ = 31,846) about gender groups. Table 3 presents tweets from our sample (we do not endorse their content).

| *Tweets with social generics* |
|---|
| *Political groups* |
| (1) "Damn right! Liberals are the real homophobes!" |
| (2) "Republicans don't care if our children and teachers are slaughtered in school." |
| (3) "Democrats glorify the killing of the unborn." |
| (4) "Democrats, Republicans 🤝 thinking Vicky Hartzler is a trash human being" |
| *Gender groups* |
| (5) "Trans people who feel the need to put down xenogenders are pawns of the cishet world order." |
| (6) "Men need extra special healthcare and full access to women's and children's safe spaces, you bigot." |
| (7) "Cishetwhite people want to be oppressed so bad but without the actual oppression part." |
| (8) "Men in white coats 🤮" |
| *Ethnic groups* |
| (9) "Black people are the best at everything." |
| (10) "White people ain't scared of nothing but racial equality." |
| (11) "Asians are more anti-Black than white people." |
| (12) "White people with dreadlocks 😒" |

**Table 3.** Selection of examples of tweets with social generics from our dataset.

In relative terms, considering proportional differences in the distribution of generic versus non-generic tweets for each of the three social categories, unlike predicted, tweets about ethnic groups had the highest proportion of tweets with generics (29.8%, $n$ = 63,042 of 211,344), followed by tweets about political groups (21.2%, $n$ = 144,789 of 684,526), and tweets about gender groups (18.2%, $n$ = 31,846 of 175,395). These differences between the tweets about political groups and the tweets about gender groups ($\chi^2$ (1) = 767.32, $p$ < 0.001, φ = 0.03, OR = 1.21, 95% CI [1.19, 1.23]) and the tweets about political groups and the tweets about ethnic groups ($\chi^2$ (1) = 6824.62, $p$ < 0.001, φ = – 0.09, OR = 0.63, 95% CI [0.62, 0.64]) were statistically significant. Hence, while the use of tweets with generics about political groups was higher than the use of tweets with generics about gender groups, against H3, it was in fact lower than the use of tweets with generics about ethnic groups (for details on the performance of the classifier on the three groups of tweets, see Supplementary Material, Figure S1).

*H4: Negatively valenced tweets with generics about political groups are more common compared to negatively valenced tweets with generics about gender or ethnic groups.* To test H4, we first conducted a $\chi^2$ test comparing the percentages of social generics tweets (about political, gender, and ethnic groups) with negative, neutral, or positive sentiment. This test, which assesses the H0 that the percentages of the three sentiment groups are equal across the three social groups, showed that there was a significant difference in sentiment across the three social groups ($\chi^2(4) = 23019.12, p < 0.001, V = 0.22$). Moreover, consistent with H4, 48.6% of all social generics tweets with negative sentiment were tweets about political groups (see Table 4).

|  | *Social category* | | | |
| --- | --- | --- | --- | --- |
| *Sentiment* | political | gender | ethnic | *Total* |
| positive | 5,027 | 8,920 | 5,786 | 19,733 |
| neutral | 23,229 | 6,987 | 11,443 | 41,659 |
| negative | 116,533 | 15,939 | 45,813 | 178,285 |
| *Total* | 144,789 | 31,846 | 63,042 | 239,677 |

**Table 4.** Sentiment across tweets with generics about social groups.

Proportionally, tweets with generics about political groups also had the highest percentage of negative tweets (80.5%), followed by tweets about ethnic groups (72.7%), and tweets about gender groups (50.1%) (see Table 4 and Figure 1). Combining neutral and positive tweets and comparing their combined total number with the number of negative tweets, tweets with generics about political groups contained significantly more negative tweets compared to tweets with generics about gender ($\chi^2(1) = 12894.84, p < 0.001, \varphi = 0.27$), with the latter set of tweets having a four times higher chance of containing neutral or positive sentiment (OR = 4.12, 95% CI [4.01, 4.22]). Similarly, tweets with generics about political groups contained also significantly more negative tweets compared to tweets with generics about ethnic groups ($\chi^2(1) = 1568.65, p < 0.001, \varphi = 0.09$), which had about two times higher chances of having neutral or positive sentiment (OR = 1.55, 95% CI [1.52, 1.59]). Tweets about ethnic groups in turn contained more negative sentiment tweets compared to tweets about gender groups ($\chi^2(1) = 4763.70, p < 0.001, \varphi = -0.22$, OR = 0.38, 95% CI [0.37, 0.39]). Hence, the number of negative tweets with generics about political groups was significantly higher than the number of negative tweets about gender or ethnic groups, supporting H4.

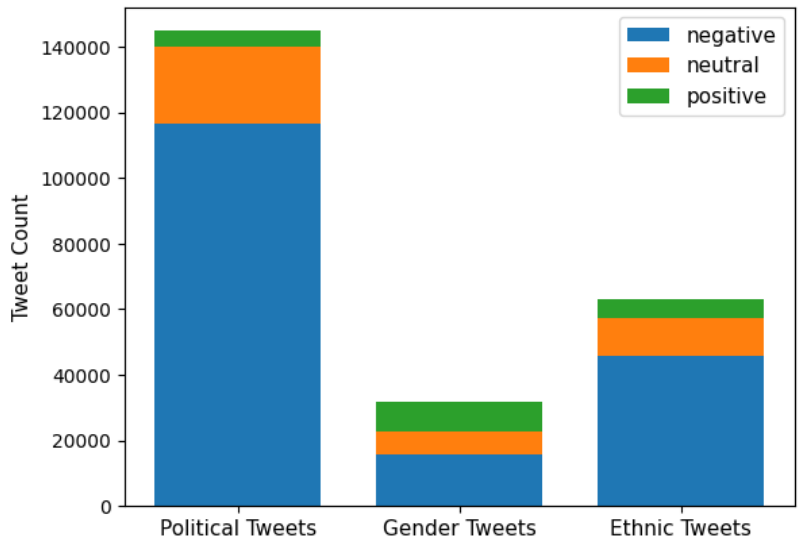

**Figure 1.** Sentiment distribution of tweets with social generics by social group.

*H5: Tweets with generics about political groups have higher "likes" and retweet impact than tweets with generics about gender or ethnic groups.* A Kruskal-Wallis *H* test of the H0 that the three groups of social generics tweets do not differ in their average mean ranks of "likes" and retweets showed that the H0 could be rejected. Both "likes" ($H(2) = 2274.554$, $p < 0.001$, $\varepsilon^2 = 0.009$) and retweet impact differed significantly between tweets with generics about political, gender, and ethnic groups ($H(2) = 1972.685$, $p < 0.001$, $\varepsilon^2 = 0.008$). But against H5, post-hoc comparisons revealed that tweets with generics about gender groups had the highest "likes" impact (*Mean rank* = 135081.00), followed by tweets with generics about political (*Mean rank* = 119214.71) and ethnic groups (*Mean rank* = 113573.24) ($p < 0.001$). The findings were different for retweets, however. Consistent with H5, tweets with generics about political groups did receive significantly more retweets (*Mean rank* = 123223.54) than tweets with generics about gender (*Mean rank* = 120172.48), and ethnic groups (*Mean rank* = 111897.25) ($p < 0.001$). Hence, for retweets but not for "likes", H5 was confirmed.

We added an explorative test (not pre-registered), examining the prediction that negatively valenced tweets with generics about political groups would receive more "likes" and retweets than negatively valenced tweets with generics about the other groups. As predicted, the three groups of negative tweets did differ between each other in "likes" impact ($H(2) = 2462.095$, $p < 0.001$, $\varepsilon^2 = 0.014$). However, post-hoc tests showed that, unlike hypothesized, negative tweets with generics about gender groups had the highest average ranks (*Mean rank* = 106794.41), followed by negative tweets with generics about political (*Mean rank* = 88525.82) and ethnic groups (*Mean rank* = 84571.73), with all differences being significant ($p < 0.001$). There was also a significant difference in the number of retweets between the three groups of negative tweets with social generics ($H(2) = 1419.981$, $p < 0.001$, $\varepsilon^2 = 0.008$). Pairwise comparisons showed that, as predicted, negatively valenced tweets with generics about political groups did receive significantly more retweets (*Mean rank* = 91289.59) than negatively valenced tweets with generics about ethnic groups (*Mean rank* = 83070.88, $p < 0.001$). Yet, unlike predicted, we could not reject the H0 that negative tweets with generics about political groups had the same average retweet impact as negative tweets with generics about gender groups (*Mean rank* = 90901.78, $p = 0.754$). That is, there was no evidence that negative tweets with generics about political groups received more retweets than negative tweets with generics about gender groups. They only differed in this respect from negative tweets with generics about ethnic groups.

## 6. General discussion

The results of this study offer several important and novel contributions to the research on generics and Twitter communication. We briefly revisit, discuss, and highlight key implications of our main findings.

[1] *Most tweets about people did not contain social generics.* As noted, it is a common view that the "language of generalization is ubiquitous in everyday conversation" (Tessler & Goodman, 2019, p. 4), that generics are people's "default" mode of generalizing (Leslie, 2012), and that (correspondingly) "generics [are] ubiquitous in everyday communication, thought, and scientific discourse" (Sterken, 2016, p. 17; Meyer et al., 2011, p. 913; Liebesman, 2011, p. 409). If these claims about the high base rate of generalizations and generics in everyday communication are correct and one adds the common suggestion that generics use may increase when writers have space constraints (DeJesus et al., 2019), then one might strongly predict that generics are also widespread in tweets about three major social groups (political, gender, ethnic groups). Given this background, our finding that only 22% of these tweets in our sample contained such generics is striking. Importantly, our tweet sample was drawn from Twitter with a query specifically targeting tweets about social groups. And while tweets about social groups might often not contain any generalizations, social generalizations (especially in political, gender, or ethnicity related posts) are often thought to be "all over social media, Twitter especially" (Edwards, 2017; Twist, 2018; Lister, 2022), with evidence suggesting that generalizing (e.g., political) outgroup related tweets receive high engagement (Rathje et al., 2021). Given these points, if anything, our query should have made it *more* likely to result in retrievals with social generics, not less. Moreover, to avoid false negatives, we tailored our criteria for identifying social generics to tweet communication (e.g., incomplete text, emojis, etc.) and made our criteria more lenient than those previously employed for generics extractions from academic texts (DeJesus et al., 2019; Peters & Lemeire, 2023). For instance, unlike these previous studies, we also included tweets with omissions or ungrammatical expressions that implied a social generic (e.g., "yt people really be bothered by everything that black people do"). That we nevertheless found social generics in only about 1/5 of all tweets about people calls into question the notion that such generics are ubiquitous in people's Twitter communication.

That said, 22% remains a large quantity. And importantly, among the many grammatical errors in social media communication, ommissions are especially common (Sihotang et al., 2021). This may have contributed to the high number of non-generic tweets because although we included incomplete tweets if they involved a generic noun phrase and at least an implied present tense predicate, many tweets contained just single words, emoticons, or links. These tweets may have constituted a significant proportion of our entire dataset, potentially accounting for our finding of a relatively low overall frequency of generics.

Nonetheless, this finding also raises doubts about frequent claims that the "use of stereotypes […] is prevalent on […] Twitter" (Taunk et al., 2022, p. 1). Stereotypes may become expressed in tweets in different ways, not only in generics. However, social generics are commonly thought to be one of their paradigmatic linguistic manifestations (Leslie, 2017; Gelman, 2021). Moreover, some previous research on (e.g., ageist) stereotypes on social media (e.g., Facebook and Twitter) did find stigmatizing group targeting language in up to about 50% of group descriptions (Levy et al., 2014; Oscar et al., 2017). Yet, if generic stereotypes about social groups were pervasively used on Twitter, social generics should have appeared more frequently in our sample of tweets than in just 1/5 of them. By revealing a comparatively low use of social generics in a large set of tweets retrieved with a query geared toward obtaining posts about social groups, our results challenge common claims about stereotyping language use on Twitter at least if one typical kind of such language (i.e., social generics) is concerned. To the extent that our results are generalizable, they may give hope because they suggest that social generics, and so language that can fuel stereotypes, may be less widely used in messages on one of the biggest social media platforms today than one might fear based on recent claims.

[2] *Tweets with social generics received more "likes" and retweets than those without.* It is often assumed that exaggerating, overgeneralizing language may draw more attention to claims, helping to underline their importance (Sumner et al., 2016; Intemann, 2022; Peters & Lemeire, 2023). Since generics are an instance of such language (DeJesus et al., 2019), our finding that tweets with social generics received both more "likes" and retweets than tweets without them supports this assumption. In doing so, our result offers a new contribution to research into the virality of tweets because even though many features of tweets (e.g., negativity, emotionality, 'fake news' status, etc.) have been linked to increased spread of content (Tsugawa & Hiroyuki, 2015; Vosoughi et al., 2018; Bellovary et al., 2021), genericity has until now not been examined in this context.

[3] *Tweets with generics about political groups were less common than tweets with generics about ethnic groups.* Recent contributions in political psychology have suggested that bias and hostility against political opponents might be more common than bias and hostility directed at gender or ethnic groups, as there seem to be weaker social norms to constrain the former tendencies (Iyengar & Westwood, 2015; Westwood et al., 2018; Iyengar et al., 2019; Peters, 2022). Some commentators have even claimed that unlike gender or ethnicity, "political identity is fair game for hatred" (Klein & Chang, 2015). Combining this with the frequent suggestion that political polarization is rife on Twitter (Hong & Kim, 2016; Urman, 2020) and the point that social generics are often manifestations of social bias and stereotypes, one might expect tweets with generics about political groups to be more common than tweets with generics about gender and ethnic groups. However, with respect to tweets about ethnic groups, we found the opposite. This

is remarkable because if political polarization were widespread on Twitter, then, since political opponents often tend to exaggerate each other's extremity (Graham et al., 2012) and generics are prime vehicles to do so (Leslie, 2017), one may think social generics are more prevalent in tweets about political groups. Still, our results remain disconcerting not least because they suggest that descriptions of ethnic groups that obscure variations and can fuel harmful ethnic stereotypes may be more common on Twitter than such descriptions about political groups despite the existence of social norms to protect members of ethnic groups.

[4] *Negatively valenced tweets with generics about political groups were more common than negatively valenced tweets with generics about gender or ethnic groups, and they received more retweets than negatively valenced tweets with generics about ethnic groups but fewer "likes".* These results add nuance to the previous finding because even though tweets with generics about political groups were less common compared to tweets about ethnic groups, they were more frequently negative in nature. This may support the notion that political animosity and bias are currently less constrained by social norms.

Indeed, our additional (explorative) result that negative tweets with generics about political groups obtained more retweets than their counterpart for ethnic groups aligns with previous research that found that political bias and discrimination were more pronounced than ethnic bias and discrimination (Iyengar & Westwood, 2015). While retweets typically express endorsement (Metaxas et al., 2015), even if users instead retweeted to directly challenge or mock a position, the increased retweets would still suggest Twitter users engaged more with, and so contributed more to the spread of, negative generic content about political groups compared to negative content related to ethnic groups. This finding is consistent with other studies reporting that negative political tweets (e.g., US Supreme Court approval of same-sex marriage) spread more than positive ones (Schöne et al., 2021). Our result adds a new perspective to this research on negative political tweets by focusing specifically on tweets with *generics* about political groups and comparing them to tweets about other groups.

It might be that people equally frequently tweeted negative generics about all three groups but that Twitter's content moderators more often removed negative tweets about ethnic groups than tweets about political groups. Twitter is not transparent about its moderation. However, we found indications that our dataset contained tweets that the platform labelled as "sensitive", suggesting that our dataset was to some extent unmoderated. Specifically, from the whole dataset, 38 political, 81 gender, and 14 ethnic group-related tweets were labelled "sensitive". But even if the higher prevalence of retweets of negative tweets with generics about political groups was due to differences in moderation, this does not undermine the view that a difference in social norms may account for the differential impact of negative tweets with generics about political groups. For the potentially different focus by Twitter's moderators may itself be a manifestation of the effects of differences in social norms.

However, we also found that negative tweets with generics about political groups had lower "likes" impact than their counterparts about gender groups. One explanation might be gleaned from studies that found that, for instance, in the US, people were increasingly displeased about partisan polarization, grew "allergic" to politics (Klar et al., 2019), and felt "exhausted when thinking about politics" (Doherty et al., 2023, p. 1). If Twitter users, too, increasingly grow tired of politics, this might account for them also decreasing their "likes" of negative tweets about political groups. Many Twitter users may nonetheless at the same time still retweet negative content about, for instance, political opponents because negativity attracts attention and Twitter algorithms reward such tweets as they enhance engagement (Rose-Stockwell, 2023). Relatedly, psychological studies found that people had a "negativity bias", i.e., they attended more to negative, potentially harmful information than to positive information (Baumeister et al., 2001; Soroka et al., 2019). Since political identity is particularly close to many individuals' self-concept (Federico et al., 2018) and threats to the self-concept are particularly likely to elicit responding (Collins et al., 2021), the asymmetry we found in "likes" and retweets of tweets with generics about political groups may be expected.

Finally, that negatively valenced tweets about gender groups (men, women, LGBTQ+, etc.) received the most "likes" suggests yet again (albeit now with respect to gender) that even if existing social norms more strongly protect against gender-related than politics-related negativity, stereotyping, or animosity, the influence of these norms on tweeting behavior may be limited. To the extent that Twitter's algorithms (indirectly or directly) promote the spread of "liked" content (Hughes, 2022), the higher "likes" impact of negative gender-related generic tweets raises the possibility that the platform contributes to the spread of people's endorsement of negative gender-related stereotypes.

## 7. Limitations

To obtain our sample, we searched only for tweets about political, gender, and ethnic groups. Tweets about other groups may contain more or fewer social generics. Additionally, we only focused on English language tweets. Moreover, our search string was geared toward Western political groups, i.e., groups from the US (e.g., liberals, conservatives, Democrats, Republicans), UK, and Europe. Our results may not hold for tweets about non-Western groups. Furthermore, while our classifier has a high accuracy (Table 2), looking at tweets that were categorized as generics, there were also false positives, and with a more stringent threshold for binary categorization than the default 0.50, fewer false positives may result. Conversely, since this default threshold is arbitrary, some researchers may view it as too high and cite it to account for our relatively low social generics count. Relatedly, our generics identification criteria may have excluded some expressions that some researchers might treat as social generics. However, as emphasized, we adopted more liberal generics identification criteria than previous studies that focused on academic texts did. Additionally, to be as transparent as possible, in the Supplementary Material, we list our criteria and flag potential tweets that might be treated as generics tweets but that we excluded. Since no established identification criteria for social generics on social media exist yet, we welcome future research exploring different (more stringent or inclusive) criteria. More generally, we have made our models and training/test data available and encourage other researchers to refine our models (e.g., by re-training, exploring other architectures, or increasing the training data).[5]

## 8. Conclusion

Generics are intriguing linguistic expressions and have received much attention in different disciplines, ranging from AI, to linguistics, philosophy, and psychology. Several philosophers have suggested that generics are people's default mode of forming generalizations about experiences and that they can, especially when they are about social groups, contribute to the spread of harmful stereotypes and polarization. This makes it important to investigate how frequently generics are being used in the now perhaps most common domain of interpersonal communication, i.e., on social media such as Twitter. However, the use and impact of generics about social groups on Twitter have until now remained unexplored. No tool for the required big data processing was available.

Here, we developed such a tool and applied it to about 1.1 million tweets to collect data for testing five hypotheses about generics on Twitter. Our findings challenge several recent claims in the philosophical and psychological literature on generics, political polarization, and intergroup animosity. While it is often suggested that generics are ubiquitous in communication, we found that Twitter communication may be an exception, as most tweets about social groups did not contain generics. Furthermore, while recent contributions in political psychology support the prediction that tweets with generics about political groups are more common on Twitter than tweets with generics about ethnic groups, our results contradict this. However, consistent with recent claims that political animosity may be less constrained by social norms than animosity against gender or ethnic groups, we did find that negatively valenced tweets with generics about political groups were more common than negatively valenced tweets with generics about gender or ethnic groups. Additionally, negatively valenced tweets with generics about political groups also received more retweets than their counterparts about ethnic groups.

Our study was only observational in nature, however, precluding causal claims. Nonetheless, we hope that our classifier and findings help stimulate discussion on the pervasiveness of social generics in communication and encourage philosophers and social scientists to explore a big data approach to the study of generics and their effects on society.

---

[5] Finally, the effects we found tended to be only small. However, focusing solely on large effects prevents a nuanced examination of complex social and psychological phenomena, which are unlikely to be explained by a few powerful predictors. Moreover, even very small effect sizes may be robust predictors and can suggest strong support for a given phenomenon if they have cumulative consequences (Cortina & Landis, 2009), or are found in "big data", including hundreds of thousand tweets (Matz et al., 2017).

**Statements and Declarations**

*Data availability.* Python notebooks, models, instructions, and datasets consisting of tweet IDs of tweets (i.e., 'dehydrated' tweets) with all labels (i.e., group, genericity scores, sentiment labels) are available here. Twitter restricts the redistribution of Twitter content to third parties. We therefore cannot publicly share all tweet contents or metrics. But our tweet datasets can be 'hydrated' (reconstructed) using tools such as *twarc*, and we are happy to share the data privately (with the relevant anonymization).

**Acknowledgements**

Many thanks to Jacob Stegenga for facilitating our collaboration via the Cambridge-LMU AI research exchange. Thanks also to Ben Yin-Chee and Olivier Lemeire for helpful feedback on earlier drafts.

**Declaration of conflicting interests**

The authors have no conflicts of interest to declare.

**Funding**

The authors have no funding to declare.

**Supplementary Material**

*(1) Twitter API search query:*

(trans OR cis OR (white men) OR (white women) OR (white people) OR (black men) OR (black woman) OR (black people) OR (asian men) OR (asian women) OR (asian people) OR (asian people) OR (hispanic men) OR (hispanic women) OR (hispanic people) OR (hispanic people) OR (african men) OR (african women) OR (african people) OR (european men) OR (european women) OR (european people) OR (american men) OR (american women) OR (american people) OR democrats OR democrat OR republicans OR republican OR liberals OR liberal OR conservatives OR conservative OR (right wing) OR (left wing) OR leftist OR rightist OR libertarians OR libertarian OR socialist OR socialists OR moderates OR moderate OR nazi OR nazis OR marxist OR marxists OR politician OR politicians OR activist OR activists OR partisan OR partisans) -has:links -has:mentions -is:retweet -is:reply -is:nullcast lang:en

---

*(2) Social generics categorization instructions*

There are no established criteria for identifying social generics in tweets. However, there are studies in which researchers manually extracted generics from scientific articles. One such study is DeJesus et al. (2019). DeJesus et al. focused on psychology papers and used the following decision tree to assess whether a line in a paper included generic language (see 2019, p. 18371, and Supplementary Information):

1. If the line is not a complete sentence, then it is uncodable.
2. If the sentence is exclusively in the past tense, then it is not generic.
3. If the sentence is not exclusively in the past tense, and if it makes a broad claim that refers to categories (e.g., "children") or abstract concepts (e.g., "parental warmth") instead of specific exemplars (e.g., "the children tested in these experiments" or "the warmth of parents in this study"), then it is generic.
A key question posed to coders was: Are the authors extending their findings generally to members of the category or instances of the phenomenon tested, or are they describing their results in terms of the specific participants or study?
4. If the sentence is generic, the next step is to decide whether the generic is bare (i.e., has no additional information to qualify or frame the results), hedged, or framed. "Generics make broad claims about a category as a whole, as distinct from individuals, without reference to frequencies, probabilities, or statistical distributions."
"Bare" is a generic sentence that is unqualified and not linked to any particular study, such as "infants make inferences about social categories" or "adolescent earthquake survivors' [sic] show increased prefrontal cortex activation to masked earthquake images as adults."
"Framed" is a generic that is unqualified, but is framed as a conclusion from the particular study that was conducted, such as "Moreover, the present study found that dysphorics show an altered behavioral response to punishment" or "We show that control separately influences perceptions of intention and causation".

"Hedged" is a generic that is qualified by a phrase, such as "These results suggest that leaders emerge because they are able to say the right things at the right time" or "Thus, sleep appears to selectively affect the brain's prediction and error detection systems," or a qualifying adverb (e.g., "perhaps") or auxiliary verb (e.g., "may"), such as "Mapping time words to perceived durations may require learning formal definitions. (DeJesus et al., 2019, p. 1837)

*Selection criteria.*

To identify generics in tweets, we partly adopted and modified DeJesus et al.'s criteria to the following list:

(1) If the sentence is exclusively in the past tense, then it is not generic.
(2) If the sentence is not exclusively in the past tense, and if it makes a broad claim that refers to categories or abstract concepts instead of specific exemplars, then it is generic.
(3) We focused only on generics about people or social groups.
(4) We grouped DeJesus et al.'s three categories of (bare, framed, hedged) generics together.
(5) As bare generics, we also included 'should' constructions (e.g., 'Democrats Should Be Less Boring').
(6) As framed generics, we also included, e.g., 'Most people think that Ks do F', and embedded generic noun phrase, e.g., 'X asks Y to help Ks', 'X is waiting for Ks to F', 'That's right people who call everything "woke" are called terfs', 'BREAKING NEWS: Democrats expand lead in US Senate').
(7) As hedged generics, we also included adverbially qualified generics (e.g., 'Ks usually, generally, consistently, often etc. do F', 'Ks probably/potentially/possibly, etc. are (or do) F')), as well as 'Ks can Y', and 'Ks would Y' generics.
(8) DeJesus et al. focused on grammatically correct, academic language (e.g., complete sentences). Applying these criteria to tweets without adaptation is problematic because tweets are known to significantly deviate from academic and personal face-to-face communication in many ways and frequently contain incorrect grammar.

To avoid underestimating the use of generics on Twitter, we first had two researchers with expertise on generics and experience with Twitter communication look at 500 example tweets to identify recurring linguistic structures in tweets that they would view as conveying generic claims even if they lacked components of grammatically correct, full sentences. We subsequently added these linguistic structures to DeJesus et al.'s criteria. They included:

(1) 'People (black, girls, etc.) be like ooh', "White ppl be like __", 'yt people really be bothered by everything that black people do' = these expressions miss 'would' or 'had', or include abbreviations

(2) 'white (black, etc.) people when a minority tells them they can't say a slur', 'Men in white coats 🤮', 'White people with dreadlocks 😐', 'Black men that can sing omg 🥺', 'white women with under cuts—it's always a 🚩', 'white people 🤝 gatekeeping natural attractions' = these expressions use social generics to refer to images or other files to 'complete the sentence'

(3) 'Black Democrats more likely than Republicans to think __', '"women who are taking hormones to be men" trans men.', 'Democrats the party of moochers', 'white people: "well i guess everything racist nowadays then huh!?" … pretty much', 'Democrats. Just stupid as fuck', 'Asian people: Smart ass hell, but dont know how to weight produce at the self checkout', 'black people so funny 😂', 'Democrat = Autocrat' = these expressions miss 'are' but clearly communicate generics

(4) 'democrat deliver communism', 'COVID Theater is the hoops Republican's are jumping through to hide the real FL COVID numbers from the public', 'straight men never likes other men's deep voices and sunoo liked it', 'white people is really protected by the damn fucking system', 'Chinese do not love black people.' = these expressions contain, e.g., incorrect lower/upper case, plural 's', genitive and/or plural mistakes, or missing (but implied by) verb components

*Other included structures:*

'Fascinated by people who can't see that homophobia and transphobia go hand in hand and somehow think trans people causing a rise in homophobia - where the fuck do they hang out?'
'nah that one tweet that was mocking how trans people be on here asking for money is hilarious😭😭😭😭'
'heres how american trans kids and minorities are being hurt by the war in ukraine'
'Realizing it's not just black people that get offended when members of our race seem to shun parts of cultures found in our group.'
'The people who really want to see the country prosper are not parading themselves as politicians. They are just change makers and I know few of them.'
'y'all give white women way too many chances at redemption when most of the time they won't even agree that they did something wrong, and yes this is about ms wilde'
'hate it when people talk bout black on black crime as if same race crime don't happen in literally every race.'
'BREAKING REPORT: Democrats BLOCKING Republican Efforts To Digitally Upload Hunter Biden's Laptop.'
'wild how white men can just. decide to sleep anywhere and not worry for their lives'
'The same thing that happened to the neanderthal is happening to modern white people. They're going extinct.'
'When politicians say 'solar and wind are cheaper than nuclear' - they are lying.'
'i hate when people including white people ask me "are you african" like bitch gtf out my face'
'Another funny thing about racism, nobody likes black people right , but mfas wana join forces when they see that whites dont like they asses either 😂.'
'Funny watching republicans idealize black football players who think they are racist terrorist nazis.'

We excluded conditional statements ('If white racists people are afraid of becoming the minority [__]'), question structures (e.g., 'Will Trudeau's Liberals choose transparency or cover up Emergencies Act ...'), and expressions without any feature ascription (e.g., "The Conservatives are about whether"), i.e., we focused only on indicative structures linking groups with properties. We also excluded any 'shoutouts' meant to directly call out a group (e.g., 'conservatives who stood against ukraine. hahahahah bye bye dumbasses', 'trans kids, hold on.').

*Other excluded structures:*

'Scruffy look on black men 😌😌😌😌'
'I need white people to learn the word excuse me .'
'WHITE PEOPLE MASALA 😭😭😭😭'
'Middle-to-upper class white women being white on TikTok is one of my favorite genres. No worries, just optimism.'
'i hate white people so much'
'Old men driving past us and staring disapprovingly because we have a "Protect Trans Kids" bumper sticker is the most Calgary thing ever.'
'White people cancelling white people over blackfishing allegations will be the funniest thing for me'

*Additional results*

For more descriptive insights into potential differences in the performance of the classifier on the three groups of tweets, we also compared the genericity scores of the groups. The genericity score between the three tweet groups differed significantly ($H(2) = 3484.242$, $p < 0.001$, $\varepsilon^2 = 0.044$). Post-hoc pairwise comparisons showed that tweets about gender groups had the highest score (*Mean rank* = 140169.58), followed by tweets about ethnic (*Mean rank* = 136418.47) and political groups (*Mean rank* = 108148.54) ($p < 0.001$) (see Figure S1).

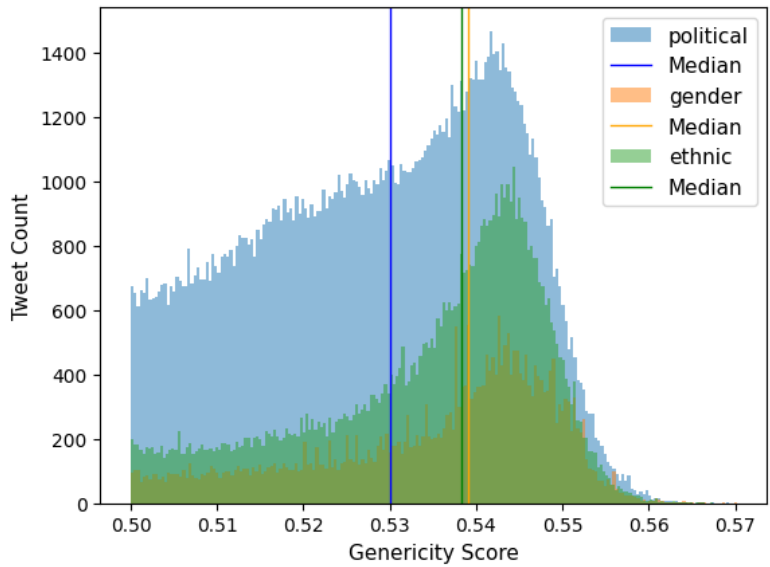

**Figure S1.** Histogram showing the distribution of tweets and median genericity scores ($\geq 0.50$) by group.

For an overview of all scores without cut-off, see Figures S2 and S3.

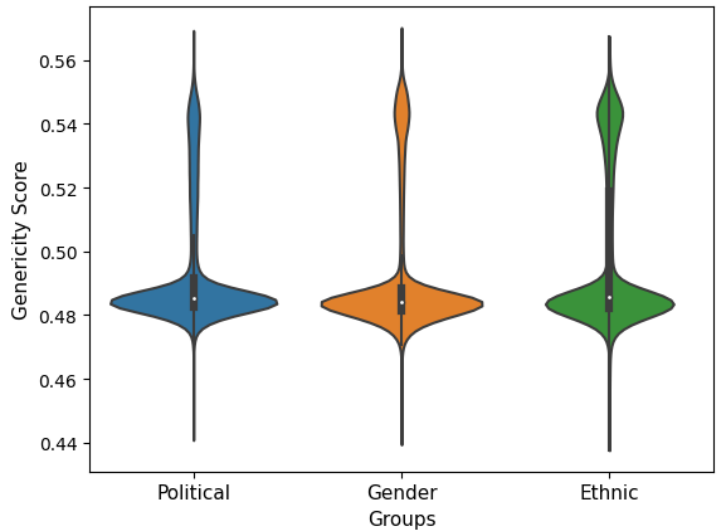

**Figure S2.** Violin plot of all tweets and their genericity score by social group, depicting median (white dot), IQR (thick black bar), and the distribution shape of the data (wider sections of the plot = higher probability that tweets take on the given value; slimmer sections = lower probability)

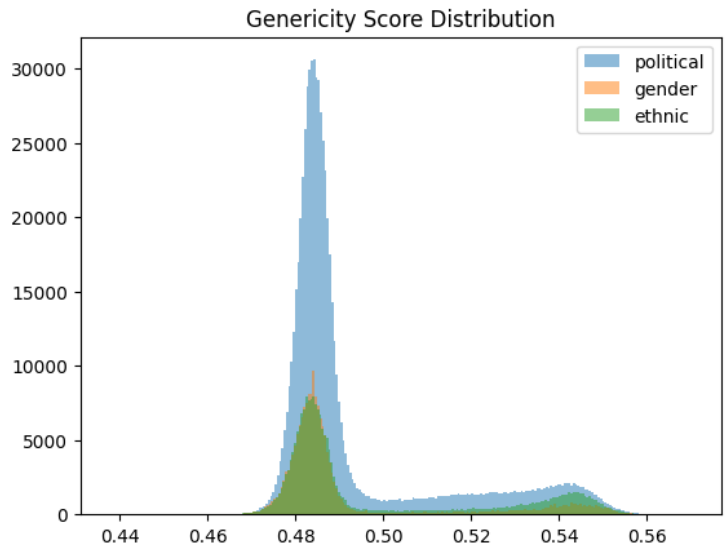

**Figure S3.** Histogram showing the total distribution of genericity scores across all tweets.